\documentclass[aps,pra,showpacs,twocolumn]{revtex4}
\usepackage{amsfonts}
\usepackage{mathrsfs}
\usepackage{amssymb}
\usepackage{amsmath}
\usepackage{graphicx}

\begin{document}

\title{Geometric phase and quantum phase transition in an inhomogeneous
periodic $XY$ spin-1/2 model}
\author{Yu-Quan Ma and Shu Chen}
\affiliation{Institute of Physics, Chinese Academy of Sciences, Beijing 100190, China}
\date{\today }

\begin{abstract}
The notion of geometric phase has been recently introduced to analyze the
quantum phase transitions of many-body systems from the geometrical
perspective. In this work, we study the geometric phase of the ground state
for an inhomogeneous period-two anisotropic $XY$ model in a transverse
field. This model encompasses a group of familiar spin models as its special
cases and shows a richer critical behavior. The exact solution is obtained
by mapping on a fermionic system through the Jordan-Wigner transformation
and constructing the relevant canonical transformation to realize the
diagonalization of the Hamiltonian coupled in the $k$-space. The results
show that there may exist more than one quantum phase transition point at
some parameter regions and these transition points correspond to the
divergence or extremum properties of the Berry curvature.
\end{abstract}

\maketitle

\section{Introduction}

Since the existence of the adiabatic geometric phase that was first revealed
in pioneer work of Berry \cite{Berry}, the concepts of geometric phases have
been extensively generalized along many directions \cite{Simon, Aharonov,
Samuel}, and now the applications of geometric phases can be found in
various physical fields \cite{Shapere,Bohm,Thouless,Morpurgo,Zanardi}.
Recently, the close relation between geometric phases and quantum phase
transitions (QPTs) has been revealed gradually \cite{Carollo,Zhu,Hamma} and
increasing interest has been drawn to the role of geometric phases in
detecting QPTs for various many-body systems \cite{chengang}. Essentially,
quantum phase transitions happened at zero temperature are characterized by
the dramatic changes in the ground-state properties of a many-body system.
Unlike classical phase transitions driven by thermal fluctuations, QPTs are
driven by pure quantum fluctuations. Traditionally, QPTs are analyzed by
resorting to notions such as the order parameter and symmetry breaking
within the Landau-Ginzburg paradigm, however, this scheme cannot give a
complete description of properties of the ground state in many-body systems.
In the past few years, a lot of efforts have been devoted to the study of
QPTs from other perspectives, such as quantum order or topological order
\cite{Wen}, entanglement measures \cite{Gu,Nielsen}, and quantum fidelity
based on the concept of quantum information \cite%
{Sun,ZanardiPRL,chenshu,Gu08}. Generally, in the vicinity of QPTs, the
changes in the ground state driven by external parameters of the Hamiltonian
will lead to an energy-level crossing or avoided energy-level crossing
between the ground state and the excited state \cite{Sachdev}, and the
features of level structures can be captured by the geometric phase of the
ground state because the features of energy-levels crossing or avoided
crossing correspond to the divergence or extremum property of the Berry
curvature. From the geometrical perspective, the geometric phase is a
reflection of the global curvature in the parameter space of the Hamiltonian.

In the present work, we shall use the geometric phase of ground state to
detect the QPTs for a period-two inhomogeneous anisotropic $XY$ spin-$\frac{1%
}{2}$ chain in a transverse field, in which the nearest-neighbor
interactions and the degree of anisotropy will take alternating parameters
between the neighbor sites \cite{Tong,Feldman,Lima,Derzhko,Tong02,Ge}. To
the best of our knowledge, the previous studies of geometric phase as a
witness of QPTs mainly concerned homogeneous spin chains for simplicity \cite%
{Carollo,Zhu,Hamma}. On the other hand, inhomogeneous systems will exhibit
rich phase diagrams and it would be interesting to investigate whether the
geometric phase is able to characterize the quantum phase transition in
these more complicate systems. So far, many methods have been introduced to
investigate the inhomogeneous spin chains in different limited conditions
\cite{Tong,Feldman,Lima,Derzhko,Tong02,Ge,Derzhko2}; however an explicit
expression of the ground state, which is necessary for the derivation of the
geometric phase, is still lacking. In our scheme, by mapping the spin
Hamiltonian on a fermionic system through the Jordan-Wigner transformation
and the Fourier transformation, we derive a general canonical transformation
to realize the diagonalization of the fermionic system Hamiltonian coupled
in the $k$- space and construct the exact expression of the ground state.
Our results show that there exist more than one critical points at some
parameter region and the critical points correspond to the divergence or
extremum property of the Berry curvature of the Hamiltonian parameter space.

\section{Model}

The system under consideration is an inhomogeneous periodic anisotropic $XY$
spin-$\frac{1}{2}$ chain \cite{Tong,Feldman,Lima,Derzhko,Tong02,Ge}, which
consists of $N$ cells with two sites in each cell, and in an external
magnetic field. Its Hamiltonian is given by
\begin{eqnarray}
\mathscr{H}=&-&\sum\limits_{l=1}^{N}[J_{a}(1+%
\gamma_{a})S_{l,a}^{x}S_{l,b}^{x} +J_{a}(1-\gamma_{a})S_{l,a}^{y}S_{l,b}^{y}
\notag \\
&+&J_{b}(1+\gamma_{b})S_{l,b}^{x}S_{l+1,a}^{x} +J_{b}(1-
\gamma_{b})S_{l,b}^{y}S_{l+1,a}^{y}  \notag \\
&+&h(S_{l,a}^{z}+S_{l,b}^{z})],  \label{spinham}
\end{eqnarray}
where $S_{l,m}^{\alpha} (\alpha=x,\ y,\ z;\ m=a,\ b)$ are the local spin
operators, $J_{m}$ is the exchange coupling, $\gamma_{m}$ is the anisotropy
in the in-plane interaction on the $m$ site in the $l$th cell and $h$ is the
external field. We assume periodic boundary conditions and choose $N$ to be
odd for convenience. This model encompasses a group of other well-known spin
models as its special cases \cite{Lieb}, such as quantum Ising model in a
transverse field for $\gamma_{a}=\gamma_{b}=1$ and $J_{a}=J_{b}$, the
transverse field $XX$ model for $\gamma_{a}=\gamma_{b}=0$ and $J_{a}=J_{b}$,
and the uniform transverse field anisotropic $XY$ model for $%
\gamma_{a}=\gamma_{b}$ and $J_{a}=J_{b}$.

In order to obtain the geometric phase in this system, we consider rotating
this model by applying a rotation of angle $\varphi $ about the $z$ axis to
each spin \cite{Carollo}, i.e., $\mathscr{D}_{z}\left( \varphi \right)
=\prod_{l=1}^{N}\exp [i\varphi (S_{l,a}^{z}+S_{l,b}^{z})]$, and then we have
$\mathscr{H}\left( \varphi \right) =\mathscr{D}_{z}^{\dagger }\left( \varphi
\right) \mathscr{H}\mathscr{D}_{z}\left( \varphi \right) $, in which $%
\mathscr{D}_{z}\left( \varphi \right) $ is the relevant rotation operator
and we have set $\hbar =1$ for simplicity. It can be verified that $%
\mathscr{H}(0)=\mathscr{H}(\pi )$, and $\mathscr{H}(\varphi )$ is $\pi $
periodic in $\varphi $ because the quadratic form about the $x$ and $y$ axes
appears symmetric in Eq. (\ref{spinham}). Considering the unitarity of the
rotation operator $U\left( \varphi \right) $ , the critical behavior and
energy spectrum of the family of Hamiltonians parametrized by $\varphi $ are
obviously $\varphi $ independent. The spin Hamiltonian can be mapped exactly
on a spinless fermion model through the Jordan-Wigner transformation
\begin{eqnarray}
S_{l,a}^{+}\! &=&\!\exp \left[ i\pi \!\!\sum_{l^{\prime
}=1}^{l-1}\!\sum_{m^{\prime }=a,b}C_{l^{\prime },m^{\prime }}^{\dag
}C_{l^{\prime },m^{\prime }}\right] \,C_{l,a}^{\dag },  \notag \\
S_{l,b}^{+}\! &=&\!\exp \left[ i\pi \!\!\left( \sum_{l^{\prime
}=1}^{l-1}\!\sum_{m^{\prime }=a,b}C_{l^{\prime },m^{\prime }}^{\dag
}C_{l^{\prime },m^{\prime }}+C_{l,a}^{\dag }C_{l,a}\right) \right]
\,C_{l,b}^{\dag },  \notag \\
&&
\end{eqnarray}%
where $S_{l,m}^{\pm }=S_{l,m}^{x}\pm iS_{l,m}^{x}$ are the spin ladder
operators and $c_{l,m}^{\dag }$ and $c_{l,m}$ are the fermion creation and
annihilation operators. The original Hamiltonian $\mathscr{H}(\varphi )$ is
transformed into
\begin{eqnarray}
\mathcal{H}(\varphi )=- &&\sum_{l=1}^{N}[\frac{J_{a}}{2}(C_{l,a}^{\dag
}C_{l,b}+C_{l,b}^{\dag }C_{l,a})+\frac{J_{b}}{2}(C_{l,b}^{\dag }C_{l+1,a}
\notag \\
&+&C_{l+1,a}^{\dag }C_{l,b})+(\frac{J_{a}\gamma _{a}}{2}e^{-i2\varphi
}C_{l,a}^{\dag }C_{l,b}^{\dag }+H.C.)  \notag \\
&+&(\frac{J_{b}\gamma _{b}}{2}e^{-i2\varphi }C_{l,b}^{\dag }C_{l+1,a}^{\dag
}+H.C.)  \notag \\
&+&h(C_{l,a}^{\dag }C_{l,a}+C_{l,b}^{\dag }C_{l,b}-1)]\ .
\end{eqnarray}%
In the fermion case, the periodic boundary conditions $S_{N+1,m}^{\alpha
}=S_{1,m}^{\alpha }$,$(\alpha =x,y,z;\ m=a,b)$ on the spin degrees of
freedom imply that $C_{N+1,m}=\exp [i\pi \sum_{l^{\prime
}=1}^{N}\sum_{m^{\prime }=a}^{b}C_{l^{\prime },m^{\prime }}^{\dag
}C_{l^{\prime },m^{\prime }}]C_{1,m}$, in which ($\sum_{l^{\prime
}=1}^{N}\sum_{m^{\prime }=a}^{b}C_{l^{\prime },m^{\prime }}^{\dag
}C_{l^{\prime },m^{\prime }}$) is just the total fermion number $N_{F}$.
Thus the boundary conditions on the fermionic system are $C_{N+1,m}=e^{i\pi
N_{F}}C_{1,m}$, and the fermionic system will obey periodic or antiperiodic
conditions depending on whether $N_{F}$ is even or odd \cite{Fradkin}.
However, the differences between the two boundary conditions are negligible
in the thermodynamic limit where the second-order QPTs occur \cite{Lieb,Zhu}%
. Without loss of generality, we assume the periodic boundary condition on
the fermionic system, which means that $N_{F}$ is always even and $%
C_{N+1,m}=C_{1,m}$. This periodic boundary condition enables us to introduce
a Fourier transformation,
\begin{eqnarray}
C_{l,a} &=&\frac{1}{\sqrt{N}}\sum_{k}e^{ikR_{la}}\,a_{k}\ ,  \notag \\
C_{l,b} &=&\frac{1}{\sqrt{N}}\sum_{k}e^{ik\left( R_{la}+a\right) }\,b_{k}
\end{eqnarray}%
to the Hamiltonian $\mathcal{H}(\varphi )$, in which $k=(2\pi /2aN)n$ and $%
n=-\frac{N-1}{2},-\frac{N-1}{2}+1,...,\frac{N-1}{2}$. Here $R_{la}$ ($%
R_{lb}=R_{la}+a$) is defined as the coordinate of site $a$ ($b$) on the $l$%
th cell in the one-dimensional lattice with the lattice parameter $2a$.
Hence, the Hamiltonian $\mathcal{H}(\varphi )$ transformed into the momentum
space is given by
\begin{eqnarray}
H_{\varphi }\!\!=\!\!\! &-&\!\!\!\sum_{k}\{h(a_{k}^{\dag }a_{k}+b_{k}^{\dag
}b_{k}-1)\!+[(\frac{J_{a}}{2}e^{ika}+\frac{J_{b}}{2}e^{-ika})a_{k}^{\dag
}b_{k}  \notag \\
&+&H.C.]-[(\frac{J_{a}\gamma _{a}}{2}e^{2i\varphi +ika}-\frac{J_{b}\gamma
_{b}}{2}e^{2i\varphi -ika})a_{-k}b_{k}  \notag \\
&+&H.C.]\}\ .
\end{eqnarray}%
This Hamiltonian has a quadratic form in fermion operators and can be
exactly diagonalized. We note that the Hamiltonian $H_{\varphi }$ can be
expressed as $H_{\varphi }\!\!=\sum_{k}(\Gamma _{k}^{\dagger }M_{k}\Gamma
_{k}+h)$ with matrix $\Gamma _{k}^{\dagger }=(a_{k}^{\dagger
},a_{-k},b_{k}^{\dag },b_{-k})$ and $M_{k}$ is a $4\times 4$ Hermitian
matrix. Therefore, we can always find a unitary transformation matrix $U$
which can be inserted in the Hamiltonian as $H_{\varphi
}\!\!=\sum_{k}(\Gamma _{k}^{\dagger }U_{k}^{\dagger
}U_{k}M_{k}U_{k}^{\dagger }U_{k}\Gamma _{k}+h)$ and then transform the
matrix $M_{k}$ into a diagonal matrix $U_{k}M_{k}U_{k}^{\dagger }$. That is
to say, the term $U_{k}\Gamma _{k}$ is equivalent to introducing the
following canonical transformation and define a set of new operators, i.e.,
\begin{eqnarray}
\gamma _{k} &=&\frac{1}{\sqrt{2}}(e^{2i\varphi }\cos \frac{\theta _{k}}{2}%
a_{k}+e^{i\delta _{k}}e^{-i\sigma _{k}}\sin \frac{\theta _{k}}{2}%
a_{-k}^{\dag }  \notag \\
&-&e^{2i\varphi }e^{i\delta _{k}}\cos \frac{\theta _{k}}{2}b_{k}+e^{i\sigma
_{k}}\sin \frac{\theta _{k}}{2}b_{-k}^{\dag })\ ,  \notag \\
\eta _{k} &=&\frac{1}{\sqrt{2}}(-e^{-i\delta _{k}}e^{i\sigma {k}}\sin \frac{%
\theta _{k}}{2}a_{k}+e^{-2i\varphi }\cos \frac{\theta _{k}}{2}a_{-k}^{\dag }
\notag \\
&+&e^{i\sigma _{k}}\sin \frac{\theta _{k}}{2}b_{k}+e^{-2i\varphi
}e^{-i\delta _{k}}e^{2i\sigma _{k}}\cos \frac{\theta _{k}}{2}b_{-k}^{\dag
})\ ,  \notag \\
\mu _{k} &=&\frac{1}{\sqrt{2}}(e^{2i\varphi }\cos \frac{\beta _{k}}{2}%
a_{k}-e^{i\delta _{k}}e^{-i\sigma _{k}}\sin \frac{\beta _{k}}{2}a_{-k}^{\dag
}  \notag \\
&+&e^{2i\varphi }e^{i\delta _{k}}\cos \frac{\beta _{k}}{2}b_{k}+e^{i\sigma
_{k}}\sin \frac{\beta _{k}}{2}b_{-k}^{\dag })\ ,  \notag \\
\nu _{k} &=&\frac{1}{\sqrt{2}}(e^{-i\delta _{k}}e^{i\sigma _{k}}\sin \frac{%
\beta _{k}}{2}a_{k}+e^{-2i\varphi }\cos \frac{\beta _{k}}{2}a_{-k}^{\dag }
\notag \\
&+&e^{i\sigma _{k}}\sin \frac{\beta _{k}}{2}b_{k}-e^{-2i\varphi }e^{-i\delta
_{k}}e^{2i\sigma _{k}}\cos \frac{\beta _{k}}{2}b_{-k}^{\dag })\ ,
\label{cantran}
\end{eqnarray}%
where
\begin{eqnarray}
\delta _{k} &=&\arg \left( J_{a}e^{ika}+J_{b}e^{-ika}\right) ,  \notag \\
\sigma _{k} &=&\arg \left( J_{a}\gamma _{a}e^{ika}-J_{b}\gamma
_{b}e^{-ika}\right) ,  \notag \\
\zeta _{k} &=&\sqrt{J_{a}^{2}+J_{b}^{2}+2J_{a}J_{b}\cos 2ka}\ ,  \notag \\
\xi _{k} &=&\sqrt{J_{a}^{2}\gamma _{a}^{2}+J_{b}^{2}\gamma
_{b}^{2}-2J_{a}J_{b}\gamma _{a}\gamma _{b}\cos 2ka}\ ,
\end{eqnarray}%
and
\begin{eqnarray}
\cos \theta _{k} &=&\frac{h-\frac{\zeta _{k}}{2}}{\sqrt{(h-\frac{\zeta _{k}}{%
2})^{2}+(\frac{\xi _{k}}{2})^{2}}}\ ,  \notag \\
\cos \beta _{k} &=&\frac{h+\frac{\zeta _{k}}{2}}{\sqrt{(h+\frac{\zeta _{k}}{2%
})^{2}+(\frac{\xi _{k}}{2})^{2}}}\ .
\end{eqnarray}%
Using the set of quasiparticle operators $\gamma _{k},\eta _{k},\mu _{k}$
and $\nu _{k}$, we can write the Hamiltonian $H_{\varphi }$ in the explicit
diagonal form as
\begin{equation}
H_{\varphi }=\sum_{q=\gamma ,\eta ,\mu ,\nu }\sum_{k}\Lambda _{q,k}\left(
q_{k}^{\dag }q_{k}-\frac{1}{2}\right) \ ,  \label{diagham}
\end{equation}%
where $\Lambda _{q,k}\ (q=\gamma ,\eta ,\mu ,\nu )$ are the eigenvalues of
the Hamiltonian matrix $M_{k}$, and now, they are the relevant quasiparticle
energy spectrums as follows:
\begin{eqnarray}
\Lambda _{\gamma k} &=&-\frac{1}{2}(h-\frac{\zeta _{k}}{2})-\frac{1}{2}\sqrt{%
(h-\frac{\zeta _{k}}{2})^{2}+(\frac{\xi _{k}}{2})^{2}}\ ,  \notag \\
\Lambda _{\eta k} &=&-\frac{1}{2}(h-\frac{\zeta _{k}}{2})+\frac{1}{2}\sqrt{%
(h-\frac{\zeta _{k}}{2})^{2}+(\frac{\xi _{k}}{2})^{2}}\ ,  \notag \\
\Lambda _{\mu k} &=&-\frac{1}{2}(h+\frac{\zeta _{k}}{2})-\frac{1}{2}\sqrt{(h+%
\frac{\zeta _{k}}{2})^{2}+(\frac{\xi _{k}}{2})^{2}}\ ,  \notag \\
\Lambda _{\nu k} &=&-\frac{1}{2}(h+\frac{\zeta _{k}}{2})+\frac{1}{2}\sqrt{(h+%
\frac{\zeta _{k}}{2})^{2}+(\frac{\xi _{k}}{2})^{2}}\ .  \label{egspre}
\end{eqnarray}%
Furthermore, it can be verified that the general canonical transformation
Eq. (\ref{cantran}) can be reduced to the familiar Bogoliubov transformation
in the case of the uniform anisotropic $XY$ model.

\section{Geometric Phase and Quantum Phase Transition}

Now, let us focus on the geometric phase of the ground state. We have
introduced the family of Hamiltonians parameterized by $\varphi$, and this
family of Hamiltonians $\mathcal{H}(\varphi)$ can be described as a result
of adiabatic rotation of the physical system. The geometric phase of the
ground state will be accumulated when the system finish a cyclic evolution,
corresponding to varying the angle $\varphi$ from 0 to $\pi$ [$\mathscr{H}%
(\varphi)$ is $\pi$ periodic in $\varphi$].

The Hamiltonian $H_{\varphi }$ in Eq. (\ref{diagham}) has been diagonalized
in the set of quasiparticle number operators, which allows us to determine
all the eigenvalues and eigenvectors. We note that the energy spectrums $%
\Lambda _{\eta k}\geq 0$, $\Lambda _{\nu k}\geq 0$ and $\Lambda _{\gamma
k}\leq 0$, $\Lambda _{\mu k}\leq 0$. The ground state, denoted as $%
|g(\varphi )\rangle $, corresponds to the state with the lowest energy,
which consists of state with no $\eta $ and $\nu $ fermions occupied but
with $\gamma $ and $\mu $ fermions occupied. Explicitly, the ground state
can be constructed as follows
\begin{equation}
\left\vert g\left( \varphi \right) \right\rangle =\mathcal{C}^{-\frac{1}{2}%
}\prod_{k>0}\left( \gamma _{-k}^{\dag }\gamma _{k}^{\dag }\mu _{-k}^{\dag
}\mu _{k}^{\dag }\eta _{-k}\eta _{k}\nu _{-k}\nu _{k}\right) |0\rangle
_{a}\otimes |0\rangle _{b},  \label{GS}
\end{equation}%
where $\mathcal{C}^{-1/2}$ is the normalized factor and $|0\rangle _{a}$ and
$|0\rangle _{b}$ are the vacuum states of the sublattices $a$ and $b$,
respectively. It is easy to check that $\eta _{k}|g(\varphi )\rangle =0$, $%
\nu _{k}|g(\varphi )\rangle =0$ and $\gamma _{k}^{\dag }|g(\varphi )\rangle
=0$, $\mu _{k}^{\dag }|g(\varphi )\rangle =0$ for all $k$. The corresponding
ground-state energy $E_{g}$ is
\begin{equation}
E_{g}=\sum_{k}\left( \Lambda _{\gamma k}+\Lambda _{\mu k}+h\right) .
\end{equation}%
The geometric phase of the ground state, denoted $\mathscr{B}_{g}$, is given
by
\begin{equation}
\mathscr{B}_{g}=\int_{0}^{\pi }\left\langle g\left( \varphi \right)
\left\vert \,i\frac{\partial }{\partial \varphi }\,\right\vert g\left(
\varphi \right) \right\rangle \ d\varphi \ .  \label{Bg}
\end{equation}%
Substituting Eq. (\ref{GS}) into Eq. (\ref{Bg}), we have
\begin{eqnarray}
&&\mathscr{B}_{g}=\frac{1}{2\mathcal{C}}\int_{0}^{\pi }\!\!\!\phantom{}%
_{a}\langle 0|\otimes \phantom{}_{b}\langle 0|\prod_{k>0}\!\left( \nu
_{k}^{\dag }\nu _{-k}^{\dag }\eta _{k}^{\dag }\eta _{-k}^{\dag }\mu _{k}\mu
_{-k}\gamma _{k}\gamma _{-k}\right)  \notag \\
&&i\frac{\partial }{\partial \varphi }\prod_{k>0}\!\left( \gamma _{-k}^{\dag
}\gamma _{k}^{\dag }\mu _{-k}^{\dag }\mu _{k}^{\dag }\eta _{-k}\eta _{k}\nu
_{-k}\nu _{k}\right) |0\rangle _{a}\otimes |0\rangle _{b}\ d\varphi .
\end{eqnarray}%
The factor of $\frac{1}{2}$ before the normalized factor $\mathcal{C}^{-1}$
is due to the repeated calculations about the $k$ and $-k$ operators.
Straightforward calculation is tedious. Nevertheless the result can be
derived concisely from the following consideration. We note that for each
term of $\gamma _{k}\frac{\partial }{\partial \varphi }\gamma _{k}^{\dag }$
and $\gamma _{-k}\frac{\partial }{\partial \varphi }\gamma _{-k}^{\dag }$ in
the integrand yield the same results of $-2i\cos ^{2}\frac{\theta _{k}}{2}$.
In the same way, the terms of $\mu _{k}\frac{\partial }{\partial \varphi }%
\mu _{k}^{\dag }$ and $\mu _{-k}\frac{\partial }{\partial \varphi }\mu
_{-k}^{\dag }$ yield the results of $-2i\cos ^{2}\frac{\beta _{k}}{2}$, the
terms of $\eta _{k}^{+}\frac{\partial }{\partial \varphi }\eta _{k}$ and $%
\eta _{-k}^{\dag }\frac{\partial }{\partial \varphi }\eta _{-k}$ yield the
results of $-2i\cos ^{2}\frac{\theta _{k}}{2}$, and the terms of $\nu
_{k}^{\dag }\frac{\partial }{\partial \varphi }\nu _{k}$ and $\nu
_{-k}^{\dag }\frac{\partial }{\partial \varphi }\nu _{-k}$ yield the results
of $-2i\cos ^{2}\frac{\beta _{k}}{2}$. Finally, the overall result is
\begin{eqnarray}
\mathscr{B}_{g} &=&\frac{i}{2}\int\nolimits_{0}^{\pi }\sum_{k>0}8(-i\cos ^{2}%
\frac{\theta _{k}}{2}-i\cos ^{2}\frac{\beta _{k}}{2})\ d\varphi  \notag \\
&=&2\pi \left[ \left( N-1\right) +\sum_{k>0}\left( \cos \theta _{k}+\cos
\beta _{k}\right) \right] \   \notag \\
&=&2\pi \sum_{k>0}\left( \cos \theta _{k}+\cos \beta _{k}\right) .
\end{eqnarray}%
To study the quantum criticality, we are interested in the properties under
the thermodynamic limit when the size of the spin lattice $N\rightarrow
\infty $. In this case, we introduce the notation of the geometric phase
density as $\beta _{g}=\lim_{N\rightarrow \infty }\mathscr{B}_{g}/N$, thus,
we have
\begin{eqnarray}
\beta _{g} &=&\lim_{N\rightarrow \infty }\frac{2\pi }{N}\sum_{k>0}\left(
\cos \theta _{k}+\cos \beta _{k}\right)  \notag \\
&=&\int_{0}^{\pi }\left( \cos \theta _{k}+\cos \beta _{k}\right) dk\ .
\end{eqnarray}%
In this case, the summation $\frac{2\pi }{N}\sum_{k>0}$ has been replaced by
the integral $\int_{0}^{\pi }dk$ with $dk=\lim_{N\rightarrow \infty }\frac{%
2\pi }{N}$. To better understand the QPTs of this inhomogeneous
periodic model and how the geometric phase of the ground state is
used as a witness to detect them,
we present numerical results for the derivative of its geometric phase $%
\partial _{h}\beta _{g}$ as a function of different parameters ($%
J_{a},J_{b},\gamma _{a},\gamma _{b}$) in the Hamiltonian.

In Fig.1, we plot it as a function of $\alpha =J_{b}/J_{a}$ and $h$
with fixed parameters $J_{a}=1$ and $\gamma _{a}=\gamma _{b}=1$,
which describes an inhomogeneous periodic Ising model in a
transverse field $h$.
\begin{figure}[t]
\begin{center}
\includegraphics[width=3.3in]{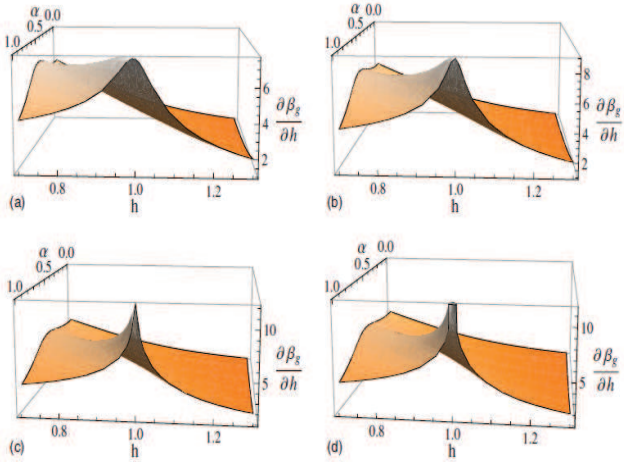}
\end{center}
\caption{(color online) (a) The derivatives of the geometric phase $d\protect%
\beta _{g}/dh$ for an inhomogeneous periodic Ising model in a transverse
field $h$, as a function of the Hamiltonian parameters $\protect\alpha %
=J_{b}/J_{a}$ and $h$, in which ($J_{a}=1,\protect\gamma _{a}=\protect\gamma %
_{b}=1$). The curves correspond to different lattice sizes $N=51$; (b) $N=101
$; (c) $N=501$; (d) $N\rightarrow \infty $.}
\end{figure}
\begin{figure}[b]
\begin{center}
\includegraphics[width=3.3in]{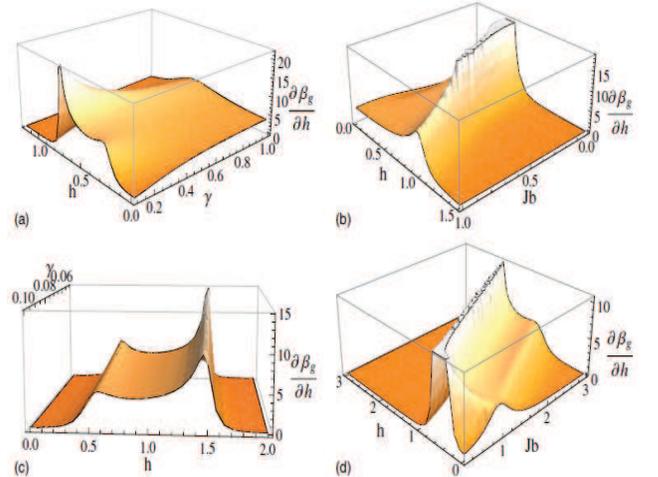}
\end{center}
\caption{(color online) (a) The derivatives of the geometric phase $d\protect%
\beta _{g}/dh$ as a function of $\protect\gamma =\protect\gamma _{a}=\protect%
\gamma _{b}$ and $h$ with the fixed parameters $J_{a}=1$, $J_{b}=0.5$ and
lattice sizes $N\rightarrow \infty $; (b) $d\protect\beta _{g}/dh$ as a
function of $J_{b}$ and $h$ with the fixed parameters $J_{a}=1$, $\protect%
\gamma _{a}=0.2$, $\protect\gamma _{b}=0.4$; (c) $d\protect\beta _{g}/dh$ as
a function of $\protect\gamma _{b}$ and $h$ with the fixed parameters $%
J_{a}=1$, $J_{b}=2$, $\protect\gamma _{a}=0.05$; (d) $d\protect\beta _{g}/dh$
as a function of $J_{b}$ and $h$ with the fixed parameters $J_{a}=1$, $%
\protect\gamma _{a}=0.2$, $\protect\gamma _{b}=0.1$.}
\end{figure}
\begin{figure}[tbh]
\includegraphics[width=3.3in]{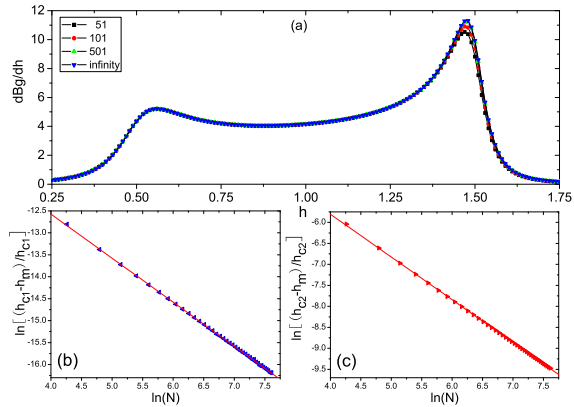}
\caption{(color online) (a) The derivatives $d\protect\beta _{g}/dh$ for the
inhomogeneous periodic $XY$ model($J_{a}=1,J_{b}=2,\protect\gamma _{a}=0.05,%
\protect\gamma _{b}=0.08$) as a function of the transverse field $h$. The
curves correspond to different lattice sizes $N=51,101,501,\infty $. (b) and
(c) The positions of the first extremum point changes and tends as $%
N^{-1.004}$ towards the first QPT point $h=0.559908$; The positions of the
second extremum changes and tends as $N^{-1.017}$ towards the second QPT
point $h=1.47561$.}
\end{figure}
As shown in Fig.1, the peak of curves for $\partial _{h}\beta
_{g}(\alpha ,h) $ becomes sharp with the increasing of the lattice
size $N$. A notable feature is that the divergence of the curve in
the thermodynamic limit only exists in the parameter region of
$J_{b}/J_{a}=1$ and $\gamma _{a}=\gamma _{b}=1$, which correspond to
the case of the uniform quantum Ising model, while in the other
parameters regions, the curves only show extremum points.

In Fig.2, we illustrate the derivative of the geometric phase of the
ground state in various cases of inhomogeneous periodic systems. An
interesting thing is that there may exist two quantum phase
transition points in some parameter regions for the inhomogeneous
period-two systems. The number of transition points and the
corresponding divergence or extremum properties of curves are
dependent on the parameters of the Hamiltonian, which is quite
different from those of the quantum Ising model and anisotropic $XY$
model in a transverse field \cite{Carollo,Zhu}. As shown in Figs.
2(a) and 2(c), the derivatives of the geometric phase only display
the extremum instead of the divergence properties even under the
thermodynamic limit condition. On the other hand, in Figs. 2(b) and
2(d), the extremum and divergence properties can coexist in some
parameter regions. In order to further understand the divergence or
extremum property of $\partial _{h}\beta _{g}(\alpha ,h)$, we choose
a section of Fig. 2(c) plotted in Fig. 3, in which the Hamiltonian
parameter takes $J_{a}=1$, $J_{b}=2$, $\gamma _{a}=0.05$ and $\gamma
_{b}=0.08$. In this case, the transition point of QPT is
characterized by the extremum point.

As shown in Fig. 3, there is no real divergence even in the thermodynamic
limit but it tends to two extremum points with the increasing of the lattice
size $N$. The transition points in the thermodynamic limit can also be
obtained by the finite-size analysis of positions of extremum points for
different size systems. Our results show that the position of the first
extremum point approaches the first QPT point $h_{c1}$ in a way of $%
h_{m}=h_{c1}(1-constN^{-1.004})$ with the transition point $h_{c1}=0.559908$
and the second one approaches as $h_{m}=h_{c2}(1-constN^{-1.017})$ with the
transition point $h_{c2}=1.47561$.

\section{Discussions and Conclusions}

As shown above, the geometry phase can be used as a detector for the more
complicated QPTs in the inhomogeneous system. This is because there exists
an intrinsic connection between the geometric phase and the energy-level
structure. Furthermore, similar connection is also reflected in the
fidelity. The relation between the fidelity and Berry phase has been
unveiled in terms of geometric tensors \cite{ZanardiPRL}. The intrinsic
relationship between the fidelity and the characterization of QPTs has also
been studied in Ref. \cite{chenshu}. For a general Hamiltonian of the
quantum many-body system undergoing QPTs given by
\begin{equation}
H(\lambda )=H_{0}+\lambda H_{1}\ ,  \label{Hlambda}
\end{equation}
where $H_{1}$ is supposed to be the driving term with $\lambda $ as the
control parameter, the second derivative of the ground-state energy can be
expressed as \cite{chenshu}
\begin{equation}
\frac{\partial ^{2}}{\partial \lambda ^{2}}E_{g}\left( \lambda \right)
=\sum_{n\neq g}\frac{2\left\vert \left\langle n(\lambda )\right\vert
H_{1}\left\vert g(\lambda )\right\rangle \right\vert ^{2}}{ E_{g}(\lambda
)-E_{n}(\lambda )}\ .  \label{2dE}
\end{equation}
Here $E_{g}$ is the ground-state energy, $n(\lambda) $ are the eigenstates
of $H(\lambda )$, and $g(\lambda)$ is the ground state. On the other hand,
the geometric phase of the system can be obtained by introducing another
parameter $\mathbf{R}$ to the Hamiltonian Eq. (\ref{Hlambda}), i.e.,
\begin{equation}
H(\mathbf{R},\lambda)=H_{0}(\mathbf{R})+\lambda H_{1}(\mathbf{R})\ .
\label{HRlambda}
\end{equation}
which is generated by a unitary transformation $H(\mathbf{R},\lambda)=%
\mathscr{U}(\mathbf{R}) H(\lambda)\mathscr{U}^{\dag}(\mathbf{R})$. Here, $%
\mathscr{U}(\mathbf{R})$ is unitary and satisfies $[\mathscr{U}(\mathbf{R}%
),H(\lambda )]\neq 0$ to ensure the nontrivial transformation. Obviously,
such a transformation keeps the energy-level structures invariant and the
critical behavior of the system is thus $\mathbf{R}$ independent. The
eigenvalues are only characterized by the parameter $\lambda$. On the other
hand, we note that the geometric phase adiabatically undergoing a closed
path $C_{\mathbf{R},\lambda}$ in the $\mathbf{R}$ space can be expressed as
\begin{equation}
\beta_g\left(C_{\mathbf{R},\lambda}\right)=-\iint_{S\left(C_{\mathbf{R}%
,\lambda}\right)} \Omega_g (\mathbf{R},\lambda)\cdot d\mathbf{S},
\end{equation}
where $\Omega_g (\mathbf{R},\lambda)$ is the Berry curvature given by
\begin{equation}  \label{Curvature}
\Omega_g (\mathbf{R},\lambda)\!=\!Im\!\sum_{n \not= g}\!\frac{\langle g_{%
\mathbf{R},\lambda}|\nabla_{\mathbf{R}} H|n_{\mathbf{R},\lambda}\rangle\!
\times\! \langle n_{\mathbf{R},\lambda}|\nabla_{\mathbf{R}} H|g_{\mathbf{R}%
,\lambda}\rangle}{(E_n(\lambda)-E_g(\lambda))^2}.
\end{equation}
From the expressions of Eqs. (\ref{2dE}) and (\ref{Curvature}), it is not
hard to find that for both of them the singularities may come from the
vanishing energy gap in the thermodynamic limit. In the inhomogeneous $XY$
model, we find that a gapless excitation occurs only when $\Lambda _{\eta
k}\rightarrow0$ or $\Lambda _{\nu k}\rightarrow0$, which demands $%
\xi_k\rightarrow0$. Hence, this condition can be achieved only in the
thermodynamic limit $N\rightarrow\infty$ and for the appropriate parameters
of the Hamiltonian, i.e., $J_a \gamma_a=J_b \gamma_b$. Apart from these
special cases, there exist no solution for $\Lambda _{\eta k}=0$ or $\Lambda
_{\nu k}=0$ and a non-zero energy gap opened. Consequently, the Berry
curvature in the thermodynamic limit only develops extremum points instead
of divergence.

In summary, we present an exact diagonalization approach for an
inhomogeneous periodic anisotropic $XY$ model in a transverse field. By
introducing a general canonical transformation, we construct an explicit
expression for the ground state, and based on this, we study the geometric
phase of the ground state and QPTs for this model. Different from the Ising
chain and anisotropic $XY$ chain in a transverse field, the inhomogeneous
periodic spin chains exhibit a richer behavior of QPTs. Our results show
that there may exist more than one phase transition point at some parameter
regions. In the language of geometric phase, detecting the QPTs of a
many-body system driven by the external parameter $\lambda$ is equivalent to
finding a path $C_{\mathbf{R},\lambda}$ in the parameter space of the
Hamiltonian, in which the Berry curvature comes to the divergence or
extremum points\cite{Hartle:1983ai}.

\begin{acknowledgments}
This work is supported by NSF of China, National Program for Basic Research
of MOST China, and programs of CAS.
\end{acknowledgments}


\begin{thebibliography}{99}
\bibitem{Berry} M. V. Berry, Proc. R. Soc. London A \textbf{392}, 45 (1984).

\bibitem{Simon} B. Simon, Phys. Rev. Lett. \textbf{51}, 2167 (1983).

\bibitem{Aharonov} Y. Aharonov and J. Anandan, Phys. Rev. Lett. \textbf{58},
1593 (1987); Y. S. Wu and H. Z. Li, Phys. Rev. B \textbf{38}, 11907 (1988).

\bibitem{Samuel} J. Samuel and R. Bhandari, Phys. Rev. Lett. \textbf{60},
2339 (1988).

\bibitem{Shapere} \textsl{Geometric Phases in Physics}, edited by A. Shapere
and F. Wilczek (World Scientific, Singapore, 1989).

\bibitem{Bohm} A. Bohm et al., \textsl{The Geometric Phase in Quantum Systems%
} (Springer, New York, 2003).

\bibitem{Thouless} D. J. Thouless, P. Ao, and Q. Niu, Phys. Rev. Lett.
\textbf{76}, 3758 (1996); D. Arovas, J. R. Schrieffer, and F. Wilczek,
\textsl{ibid.}, \textbf{53}, 722 (1984); R. Resta, Rev. Mod. Phys. \textbf{66%
}, 899 (1994).

\bibitem{Morpurgo} A. F. Morpurgo et al., Phys. Rev. Lett. \textbf{80}, 1050
(1998); S. L. Zhu and Z. D. Wang, \textsl{ibid}., \textbf{85}, 1076 (2000).

\bibitem{Zanardi} P. Zanardi and M. Rasetti, Phys. Lett. A \textbf{264}, 94
(1999); J. Pachos, P. Zanardi and M. Rasetti, Phys. Rev. A \textbf{61},
010305(R) (1999); J. A. Jones et al., Nature \textbf{403}, 869 (2000); L. M.
Duan, J. I. Cirac, and P. Zoller, Science \textbf{292}, 1695 (2001); S. L.
Zhu and Z. D. Wang, Phys. Rev. Lett. \textbf{89}, 097902 (2002); \textsl{%
ibid.},\textbf{91} 187902 (2003).

\bibitem{Carollo} A. C. M. Carollo and J. K. Pachos, Phys. Rev. Lett.
\textbf{95}, 157203 (2005).

\bibitem{Zhu} S. L. Zhu, Phys. Rev. Lett. \textbf{96}, 077206 (2006); S. L.
Zhu, Int. J. Mod. Phys. B \textbf{22}, 561 (2008).

\bibitem{Hamma} A. Hamma, quant-ph/0602091 (2006).

\bibitem{chengang} G. Chen, J. Li, and J.-Q. Liang, Phys. Rev. A \textbf{74}%
, 054101 (2006); X. X. Yi and W. Wang, \textsl{ibid}. \textbf{75}, 032103
(2007); H. T. Cui and J. Yi, \textsl{ibid}. \textbf{78}, 022101 (2008); A.
I. Nesterov and S. G. Ovchinnikov, Phys. Rev. E \textbf{78}, 015202 (R)
(2008).

\bibitem{Wen} X. G. Wen, Phys. Rev. B \textbf{40}, 007387 (1989); X. G. Wen
and Q. Niu, \textsl{ibid}. \textbf{41}, 009377 (1990); X. G. Wen, \textsl{%
Quantum Field Theory of Many-body Systems} (Oxford University Press, Oxford,
2004).

\bibitem{Nielsen} T. J. Osborne and M. A. Nielsen, Phys. Rev. A \textbf{66},
032110 (2002); A. Osterloh L. Amico, G. Falci, and R. Fazio, Nature
(London), \textbf{416}, 608 (2002); G. Vidal J. I. Latorre, E. Rico, and A.
Kitaev Phys. Rev. Lett.\textbf{90}, 227902 (2003).

\bibitem{Gu} S. J. Gu, S. S. Deng, Y. Q. Li, and H. Q. Lin, Phys. Rev. Lett.
\textbf{93}, 086402 (2004); Y. Chen, P. Zanardi, Z. D. Wang, and F. C.
Zhang, New J. Phys. \textbf{8}, 97 (2006).

\bibitem{Sun} H. T. Quan, Z. Song, X. F. Liu, P. Zanardi, and C. P. Sun,
Phys. Rev. Lett. \textbf{96}, 140604 (2006); P. Zanardi and N. Paunkovic,
Phys. Rev. E \textbf{74}, 031123 (2006).

\bibitem{ZanardiPRL} L. Campos Venuti and P. Zanardi, Phys. Rev. Lett.
\textbf{99}, 095701 (2007).

\bibitem{chenshu} S. Chen, L. Wang, Y. Hao, and Y. Wang, Phys. Rev. A
\textbf{77}, 032111 (2008).

\bibitem{Gu08} S. J. Gu, H. M. Kwok, W. Q. Ning and H. Q. Lin, Phys. Rev. B
\textbf{77}, 245109 (2008); W. L. You, Y. W. Li, and S. J. Gu, Phys. Rev. E
\textbf{76}, 022101 (2007); S. Chen, L. Wang, S. J. Gu and Y. Wang, Phys.
Rev. E \textbf{76}, 061108 (2007).

\bibitem{Sachdev} S. Sachdev, \textsl{Quantum Phase Transitions}, (Cambridge
University Press, 1999).

\bibitem{Tong} P. Tong and M. Zhong, Physica B \textbf{304}, 91 (2001); P.
Tong and X. Liu, Phys. Rev. Lett. \textbf{97}, 017201 (2006).

\bibitem{Feldman} K. E. Feldman, J. Phys. A \textbf{39}, 1039 (2006).

\bibitem{Lima} J. P. de Lima, L. L. Goncalves, and T. F. A. Alves, Phys.
Rev. B \textbf{75}, 214406 (2007).

\bibitem{Derzhko} O. Derzhko, J. Phys. A \textbf{33}, 8627 (2000); O.
Derzhko, J. Richter, T. Krokhmalskii, and O. Zaburannyi, Phys. Rev. B
\textbf{66}, 144401 (2002).

\bibitem{Tong02} P. Tong and M. Zhong, Phys. Rev. B \textbf{65}, 064421
(2002).

\bibitem{Ge} M. G. Hu, K. Xue, and M. L. Ge, Phys. Rev. A \textbf{78},
052324 (2008).

\bibitem{Derzhko2} O. Derzhko and J. Richter, Phys. Rev. B \textbf{55},
14298 (1997).

\bibitem{Lieb} E. Lieb, T. Schultz, and D. Mattis, Ann. Phys. \textbf{16},
407 (1961); E. Barouch and B. McCoy, Phys. Rev. A \textbf{3}, 786 (1971); P.
Pfeuty, Ann. Phys. \textbf{57}, 79 (1970).

\bibitem{Fradkin} E. Fradkin, \textsl{Field Theories of Condensed Matter
Systems} (Addison-Wesley, MA, 1991).


\bibitem{Hartle:1983ai} J.~B.~Hartle and S.~W.~Hawking,
Phys.\ Rev.\ D \textbf{28}, 2960 (1983). 
\end{thebibliography}
\end{document}